\documentstyle[12pt,epsfig,rotating]{article}
\def\pge{\pagestyle{empty}} \def\pgn{\pagestyle{plain}}
\def\bsls35{\baselineskip 0.35in}
     
\def\spg{\setcounter{page}} 
\def\bd{
\begin{document}} \def\ed{\end{document}}
\def\bmp{\begin{minipage}} \def\emp{\end{minipage}}
\def\bcc{\begin{center}} \def\ecc{\end{center}}     \def\npg{\newpage}
\def\beq{\begin{equation}} \def\eeq{\end{equation}} \def\hph{\hphantom}
\def\be{\begin{equation}} \def\ee{\end{equation}} \def\r#1{$^{[#1]}$}
\def\n{\noindent} \def\ni{\noindent} \def\pa{\parindent} 
\def\hs{\hskip} \def\vs{\vskip} \def\hf{\hfill} \def\ej{\vfill\eject} 
\def\cl{\centerline} \def\ob{\obeylines}  \def\ls{\leftskip}
\def\underbar#1{$\setbox0=\hbox{#1} \dp0=1.5pt \mathsurround=0pt
   \underline{\box0}$}   \def\ub{\underbar}    \def\ul{\underline} 
\def\f{\left} \def\g{\right} \def\e{{\rm e}} \def\o{\over} 
\def\vf{\varphi} \def\pl{\partial} \def\cov{{\rm cov}} \def\ch{{\rm ch}}
\def\la{\langle} \def\ra{\rangle} \def\EE{e$^+$e$^-$}
\def\bitz{\begin{itemize}} \def\eitz{\end{itemize}}
\def\btbl{\begin{tabular}} \def\etbl{\end{tabular}}
\def\btbb{\begin{tabbing}} \def\etbb{\end{tabbing}}
\def\beqar{\begin{eqnarray}} \def\eeqar{\end{eqnarray}}
\def\\{\hfill\break} \def\dit{\item{-}} \def\i{\item} 
\def\bbb{\begin{thebibliography}{9} \itemsep=-1mm}
\def\ebb{\end{thebibliography}} \def\bb{\bibitem}
\def\bpic{\begin{picture}(260,240)} \def\epic{\end{picture}}
\def\akgt{\noindent{\bf Acknowledgements}}
\def\fgn{\noindent{\bf\large\bf Figure captions}}
\def\pt{{p_{\rm t}}} \def\vf{\varphi} 
\def\yct{y_{\rm cut}} \def\kt{k_{\rm t}}
\newcommand{\QCD}{{\scshape qcd}} \newcommand{\NFM}{{\scshape nfm}}
\bd
\pge

\null{}\vskip -1.2cm
\hskip12cm{\bf HZPP-0006}
\vskip-0.2cm

\hskip12cm Sept. 5, 2000

\vskip1cm

\cl{\Large A Monte Carlo Study of the Dynamical-Flucautation }
\vskip0.4cm

\cl{\Large Property of the Hadronic System Inside Jets\footnote{
Supported in part by the National Natural Science Foundation of China 
(NSFC) under Project 19975021 .}}

\vskip1.2cm

\cl{\large Liu Lianshou,  \ \  Chen Gang\footnote{Permanent address: Jingzhou
Teacher's College, Hubei 434100 China}  \ \ and \ \ Fu Jinghua}

{\small Institute of Particle Physics, Huazhong Normal University,
Wuhan 430079 China}
\date{ }

\vskip0.5cm

\begin{center}
\begin{minipage}{125mm}
\vskip 0.5in
\begin{center}{\Large ABSTRACT}\end{center}
{\hskip0.6cm
A study of the dynamical fluctuation property of jets is carried out
using Monte Carlo method. The results suggest that, unlike the
average properties of the hadronic system inside jets,
the anisotropy of dynamical fluctuations in these systems
changes abruptly with the variation of the cut parameter $\yct$. 
A transition point exists, where the dynamical fluctuations in the
hadronic system inside jet behave like those in soft hadronic collisions,
i.e. being circular in the transverse plan with repect to dynamical
fluctuations. This finding obtained from Jetset and Herwig Monte Carlo
is encouraged to be checked by experiments.}
\end{minipage}
\end{center}
\vs0.8cm

{\large PACS number: 13.85 Hd

\vs0.5cm
\ni

\hskip1.8cm
\ processes}

\npg \pgn \spg{2}

The presently most promissing theory of strong interaction
--- Quantum Chromo-Dynamics (\QCD) has the special property of both asymptotic
freedom and colour confinement. For this reason, in any process, even though 
the energy scale, $Q^2$, is large enough for perturbative \QCD\ (p\QCD) 
to be applicable, there must be a non-perturbative hadronization phase 
before the final state particles can be observed. Therefore, the transition 
or interplay between hard and soft processes is a very important problem.

Theoretically, the transition between perturbative and non-perturbative 
\QCD\ is at a scale $Q_0 \sim$ 1--2 GeV. Experimentally, the transition 
between hard and soft processes is determined by the identification of
jets through some jet-finding processes, e.g.  Jade~\cite{Jade} or 
Durham~\cite{Durham} algorithm. In these processes there is a parameter 
--- $\yct$, which, in case of Durham algorithm, is essentially the relative 
transverse momentum $\kt$ squared~\cite{Dokshitzer},
\be 
\kt = \sqrt{\yct} \cdot \sqrt s,
\ee
where $\sqrt s$ is the center-of-mass energy of the collision.
From the experimental point of view, $\kt$ can be taken as the transition
scale between hard and soft. Its value depends on the definition of ``jet''.

Historically, the discovery in 1975~\cite{Hanson} of a two-jet structure 
in \EE\ annihilation at c.m. energies $\geq 6$ GeV has been taken
as an experimental confirmation of the parton model\cite{JELIS}, and the 
observation in 1979 of a third jet in e$^+$e$^-$ collisions at 17 -- 30 GeV 
has been recognised as the first experimental evidence of 
gluon~\cite{Brandelik} -- 
\cite{Bartel}.  These jets, being directly observable in experiments
as ``jets of particles'', will be called in the following
as ``visible jets''.  The aim of this paper is to find out the scale 
corresponding to these visible jets and discuss its meaning.

For this purpose, let us remind that
the qualitative difference between the typical soft process ---  moderate
energy hadron-hadron collisions and the typical hard process --- high 
energy e$^+$e$^-$ collisions can be observed most clearly in the property 
of dynamical fluctuations therein.  It is found recently~\cite{FFLPRD}
that the dynamical fluctuations in the hadronic systems from
these two processes are qualitatively different --- the former is
anisotropic in the longitudinal-transvere plane and isotropic in the
transverse planes while the latter is isotropic in three dimensional phase
space. This observation inspired us to think that the dynamical-fluctuation 
property may provide a hint for the determination of the scale of the
appearance of visible jets.

The dynamical fluctuations can be characterized as usual by the anomalous 
scaling of normalized factorial moments (\NFM)~\cite{BP}:
\beqar   
  F_q(M)&=&{\frac {1}{M}}\sum\limits_{m=1}^{M}{{\langle n_m(n_m-1)
     \cdots (n_m-q+1)\rangle }\over {{\langle n_m \rangle}^q}}\\ \nonumber
   &\propto& (M)^{\phi_q}\ \  \quad \quad (M\to \infty) \ \ ,
\eeqar
where a region $\Delta$ in 1-, 2- or 3-dimensional phase space is
divided into $M$ cells, $n_m$  is the multiplicity in the $m$th cell,
and $\langle\cdots\rangle$ denotes vertically averaging over the event
sample. 

Note that when the fluctuations exist in higher-dimensional (2-D
or 3-D) space, the projection effect~\cite{Ochs} will cause the 
second-order 1-D \NFM\ goes
to saturation according to the rule\footnote{In order to elliminate the 
influence of momentum conservation~\cite{MMCN}, the first few points 
($M=1,2$ or 3) should be omitted when fitting the data to Eq.(3).}:
\be   
 F_2^{(a)}(M_a) = A_a-B_a M_a^{-\gamma_a}, \ \ 
\ee
where $a=1,2,3$ denotes the different 1-D variables. The parameter $\gamma_a$
describes the rate of going to saturation of the \NFM\ in direction $a$ and is 
the most
important characteristic for the higher-dimensional dynamical fluctuations.
If $\gamma_a = \gamma_b$ the fluctuations are isotropic in the $a,b$
plane; while when $\gamma_a \neq \gamma_b$ the fluctuations are anisotropic 
in this plane. The degree of anisotropy is characterized by the Hurst
exponent $H_{ab}$, which can be obtained from the values of $\gamma_a$
and $\gamma_b$ as~\cite{ZGKX}
\be   
 H_{ab} = {1+\gamma_b\over 1+\gamma_a}.
\ee
The dynamical fluctuations are isotropic when  $H_{ab} = 1$, and anisotropic 
when $H_{ab} \neq 1$.

For the 250 GeV/$c$ $\pi$(K)-p collisions from NA22 the Hurst 
exponents are found to be~\cite{NA22}:
\be 
H_{\pt\vf}=0.99 \pm 0.01, \ \ 
 H_{y\pt}=0.48 \pm 0.06, \ \ H_{y\vf}=0.47 \pm 0.06,  
\ee
which means that the dynamical fluctuations in this moderate energy
hadron-hadron collisions are isotropic in the
transverse plane and anisotropic in the longitudinal-transvere planes.
This is what should be~\cite{WLprl}, because there is almost no hard 
collisions at this energy and the direction of motion of the incident 
hadrons (longitudinal direction) should be previleged. Note that the 
special role of longitudinal direction in these soft processes is present
both in the magnitude of average momentum and in the dynamical fluctuations 
in phase space.

In high energy e$^+$e$^-$ collisions, the longitudinal direction is chosen 
along the thrust axis, which is the direction of motion of the primary
quark-antiquark pair. Since this pair of quark and antiquark move back to back
with very high momenta, the magnitude of average momentum of final state 
hadrons is also anisotropic due to momentum conservation.  
However, the dynamical fluctuations in this case come from 
the QCD branching of partons~\cite{Vineziano}, which is isotropic
in nature. Therfore, although the momentum distribution still has 
an elongated shape, the dynamical fluctuations in this case should be 
isotropic in 3-D phase space.

A Monte Carlo study for e$^+$e$^-$ collisions at 91.2 GeV confirms this
assertion~\cite{FFLPRD}. The dynamical fluctuations are 
approximately isotropic in the 
3-D phase space, the corresponding Hurst exponents being 
\be 
H_{\pt\vf}=1.18 \pm 0.03, \ \ 
 H_{y\pt}=0.95 \pm 0.02, \ \  H_{y\vf}=1.11 \pm 0.02. 
\ee
The present available experimental data for e$^+$e$^-$ collisions at 91.2 GeV 
also show isotropic dynamical fluctuations in 3-D~\cite{DELPHI}.

Now we apply this technique to the ``2-jet'' sub-sample of \EE collision
obtained from a 
certain, e.g. Durham, jet-algorithm with some definite value of $\yct$. 
Doing the analysis for different values of $\yct$,
the dependence of dynamical-fluctuation property of the ``2-jet'' 
sample on the value of $\yct$ can be investigated.

Two event samples are constructed from Jetset7.4 and Herwig5.9 generators,
respectively, each consists of 400 000 \EE\ collision events at c.m.
energy 91.2 GeV.  The results of 3-D $F_2$ for the full samples
are shown in Fig.1 and the variation of $\gamma$'s of the 2-jet sample
with $\yct$ ($\kt$) are shown in Fig's 2. 

\begin{center}
\begin{picture}(250,450)
\put(17,230)
{
{\epsfig{file=fg3d.epsi,width=180pt,height=228pt}}
}
\put(-50,-44)
{
{\epsfig{file=fgamma.epsi,width=350pt,height=260pt}}
}
\end{picture}
\end{center}

\vskip-8.65cm
\cl{{\small Fig.1 \ The log-log plot of 3-D \NFM\ of the full event sample
as function of partition number $M$}}

\vskip9cm

\hskip4cm ($a$) \hskip6cm ($b$)

\cl{{\small Fig.2 \ The parameter $\gamma$ of 2-jet sample as function 
of $\yct$ ($\kt$)}}

\newpage

It can be seen from Fig.1 that after neglecting the first point to eliminate
the influence of momentum conservation~\cite{MMCN} the results from both
Jetset and Herwig fit very well to straight lines with only slightly different
slope. This means that the results from these two generators are 
qualitatively the same, showing that the full sample is self-similar 
(isotropic) fratal. This is just as expected~\cite{FFLPRD}. 
Quantitatively, they have slightly different fractal dimension.

The variation of the three $\gamma$'s of ``2-jet'' sample with the parameter 
$\yct$ ($\kt$), plotted in Fig.'s~2, show an interesting pattern. When
$\yct$ ($\kt$) is very small, the three  $\gamma$'s are separate. As the
increasing of $\yct$ ($\kt$), $\gamma_\pt$ and $\gamma_\vf$ approach each
other and cross over each other sharply at a certain point. After that the
three $\gamma$'s approach to a common value due to the fact that when 
$\yct$ is very large the ``2-jet'' sample coincides with the full sample
and the dynamical fluctuations in the full sample is isotropic, cf.
Eq.(6).

We will call the point where $\gamma_\pt$ crosses $\gamma_\vf$
as {\em transition point}. It has the unique property: $\gamma_\pt = 
\gamma_\vf \neq \gamma_y$, i.e. the jets at this point are circular in
the transverse plan with respect to dynamical fluctuations. 
These jets will, therefore,  be called {\em circular jets}.

The above-mentioned results are qualitatively the same for the two 
event generators, cf. Fig.2 ($a$) and ($b$), only the $\yct$ ($\kt$) values 
at the transition point are somewhat different.
It is $\yct\approx 0.0048$ ($\kt \approx 6.3$ GeV) for Jetset and 
$\yct\approx 0.0022$ ($\kt \approx 4.3$ GeV) for Herwig.  
The values of $\gamma$'s and the corresponding Hurst exponents $H$, 
cut parameter $\yct$ and relative transverse momentum $\kt$ at the 
transition point are listed in Table I.

\vs0.5cm
\cl{Table I \ \ Parameters $\gamma$, Hurst exponents $H$, cut-parameters
$\yct$ and $\kt$ at the transition point}

\vskip0.4cm

{\small 
\def\bcc{\begin{center}} \def\ecc{\end{center}}
\def\btbl{\begin{tabular}} \def\etbl{\end{tabular}} 
\def\fsz{\footnotesize}
\bcc\btbl{|c|c|c|c|c|c|c|c|c|c|}\hline
Generator {\fsz (GeV)} & $\yct^{\footnotesize{\rm (Durham)}}$  &
$\gamma_y$ & $\gamma_\pt$ & $\gamma_\vf$ &
$H_{y\pt}$ & $H_{y\vf}$ & $H_{\pt\vf}$ &
$\yct$ & $\kt$ GeV \\ \hline
Jetset7.4 & 0.0048 & 1.074 & 0.514 & 0.461 & 0.73 & 0.70 &  0.96
& 0.0048 &  6.32 \\
& {\fsz $\pm$0.0007}  & {\fsz $\pm$0.037} & {\fsz $\pm$0.080}
     & {\fsz $\pm$0.021} & {\fsz $\pm$0.06}
     & {\fsz $\pm$0.06} & {\fsz $\pm$0.10} 
& {\fsz $\pm$0.0007} & {\fsz $\pm$0.03} \\ \hline
Herwig5.9 & 0.0022 & 1.237 & 0.633 & 0.637 & 0.73 & 0.73 &  1.00
& 0.0022 & 4.28 \\
& {\fsz $\pm$0.0008}  & {\fsz $\pm$0.066} & {\fsz $\pm$0.064}
     & {\fsz $\pm$0.051} & {\fsz $\pm$0.05}
     & {\fsz $\pm$0.05} & {\fsz $\pm$0.07} 
& {\fsz $\pm$0.0008} & {\fsz $\pm$0.02} \\ \hline
\etbl\ecc}
\vskip0.2cm

It is natural to ask the question: Is there any relation between the
{\em circular jets} determined by the condition $\gamma_\pt = \gamma_\vf 
\neq \gamma_y$ and the {\em visible jets} directly observable in experiments
as ``jets of particles''?

In order to answer this question, we 
plot in Fig.'s 3 the ratios $R_2$ and $R_3$ of 
``2-jet'' and ``3-jet'' events as functions of the  relative transverse 
momentum $\kt$ at different c.m. energies. 

\begin{center}
\begin{picture}(250,450)
\put(-144,107)        
{
{\epsfig{file=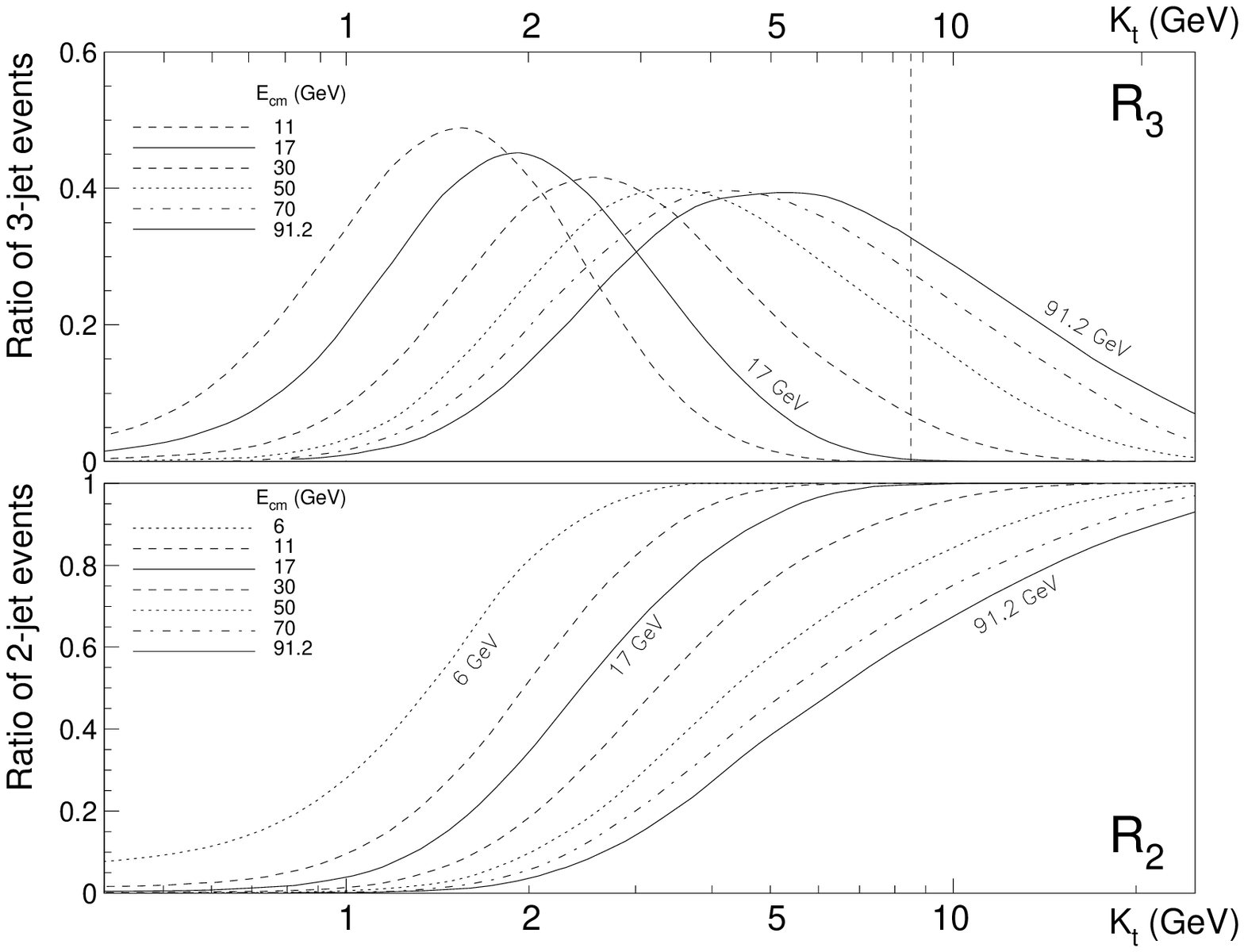,width=290.0pt,height=400pt}}
}
\put(102,107)
{
{\epsfig{file=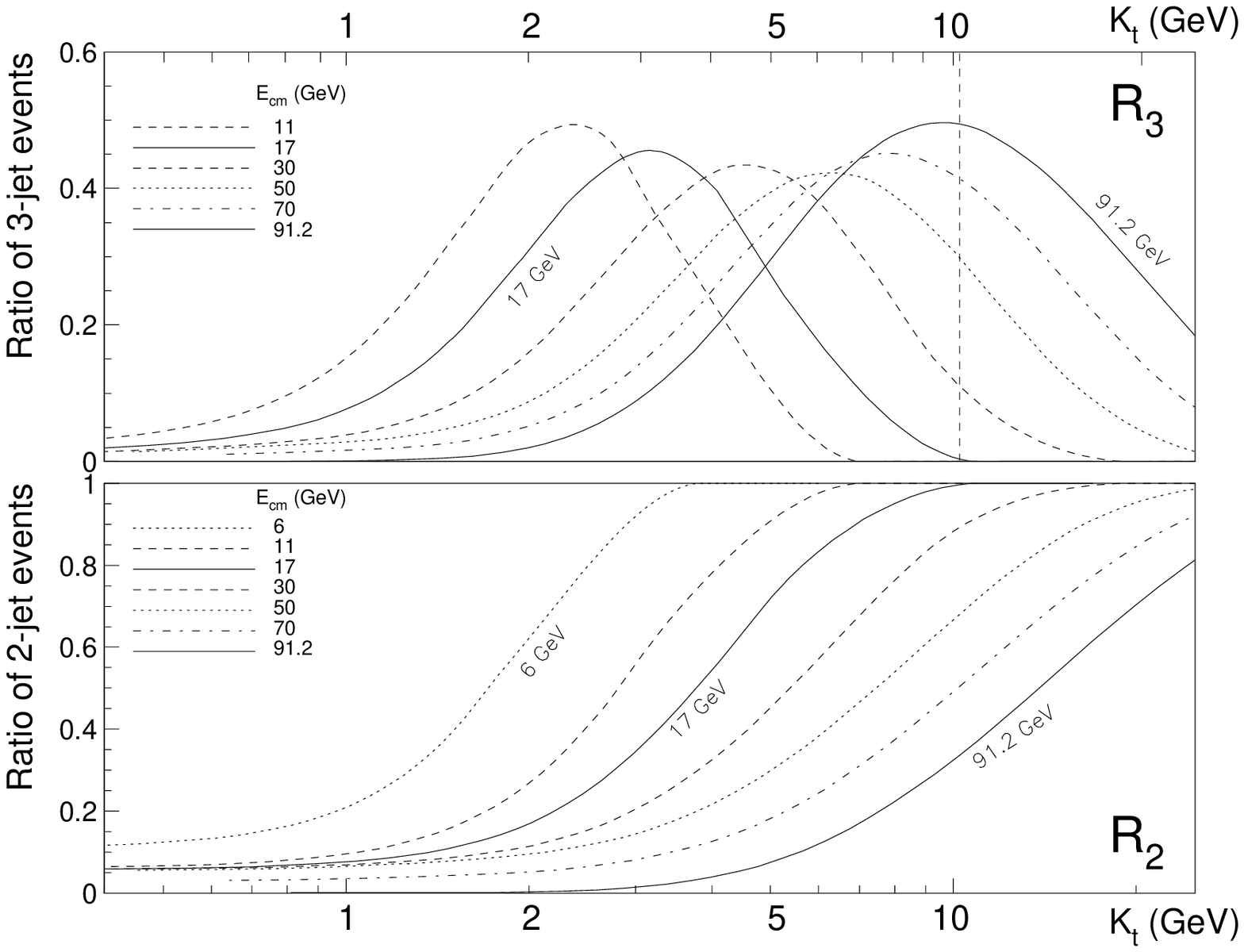,width=290.0pt,height=400pt}}
}
\end{picture}
\end{center}

\vskip-9.5cm

\hskip3.5cm ($a$) \hskip7.6cm ($b$)
 \vskip0.2cm

\hskip2cm{\small Fig.3 \ The ratios $R_3$ and $R_2$ of 3- and 2-jet events as
functions of $\kt$ 

\hskip3.0cm at different c.m. energies, ($a$) from Jetset7.4; 
($b$) from Herwig5.9}

\vskip0.5cm
Let us consider the point where a third jet starts to appear. 
Historically, a third visible jet was firstly observed in \EE\ collisions 
at c.m. energy 17 GeV. It can be seen from Fig.3 that, for $\sqrt s=$ 17 GeV,
$R_3$ starts to appear at around $\kt=$ 8--10 GeV, cf. the dashed vertical
lines in Fig.'s 3. This value of $\kt$ is consistent with 
the $\kt$ value (4.3--6.3 GeV) of a circular jet within a factor of 2, 
cf. Table I.
Thus we see that the {\em circular jet}, defined as a kind of jet
circular in the trnasverse plan with respect to dynamical fluctuations,
and the {\em visible jet}, defined as a kind of jet directly 
observable in experiments as ``jets of particles'', have about the same 
scale --- $\kt\sim$ 5--10 GeV.

This scale is to be compared with the scale $\kt\sim$ 1--2 GeV,
which is the scale for the transition between perturbative and 
non-perturbative. It is interesting also to see what happens in
the results of jet-algorithm at the latter scale. 

It can be seen from Fig.3$a$ (Jetset7.4) that, at this scale 
($\kt\sim$ 1--2 GeV) 
the ratio $R_2$ of ``2-jet'' events tends to vanish 
almost independent of energy, provided the latter is not too low.
This can be explained as the following.
Consider, for example, an event with only two hard partons, having no 
perturbative branching at all.  Even in this case, the two
partons will still undergo non-perturbative hadronization to produce
final-state particles. If the $\kt$ is chosen to be less than 1--2 GeV then 
the non-perturbative hadronization with small transverse momentum will
also be considered as the production of new ``jets'' and this 
``should-be'' 2-jet event will be taken as 
a ``multi-jet'' (more than two jets) ones too. This means that, when  
$\kt <$ 1--2 GeV, events with small transverse momentum will also
become ``multi-jet'' ones, and $R_2$ vanishes. 

However, even when $\kt <$ 1--2 GeV, 
a few 2-jet events may still survive if the hadronization 
is alomst colinear. This effect becomes observable 
when the energy is very low, see, e.g., the 
$R_2$ curve for $\sqrt s=6$ GeV in Fig.3$a$.

Similar picture holds also for the results from Herwig5.9, cf. Fig.3$b$,
but the almost-colinear hadronization appears earlier. 

Before closing the paper, let us give some comments on the physical picture
behind the above-mentioned two scales. A circular (or visible) jet is
originated from a hard parton. The production of this parton is a 
hard process. Its evolusion              
into final state particles includes a perturbative branching and subsequent 
hadronization. The hadronization is a soft process. The perturbative 
branching (sometimes called parton shower) 
in between the hard production and soft hadronization
connects these two processes.
The isotropic property of dynamical fluctuations provides a criterion for
the discrimination of the hard production of circular jets and the
parton shower inside these jets.

In this paper we found through Monte Carlo study that 
unlike the average properties of the hadronic systems inside jets,
the anisotropy of dynamical fluctuations in these systems
changes abruptly with the variation of the cut parameter $\yct$ ($\kt$). 
A transition point exists, where the dynamical fluctuations in the
hadronic system inside jets behave like those in soft hadronic collisions,
i.e. being circular in the transverse plan with repect to dynamical
fluctuations. 
The scale of these {\em circular jets} \hskip2pt is about $\kt\sim$ 5--10 GeV
in contrast to the scale of p\QCD\, which is $Q_0\sim$ 1--2 GeV.
It is encouraged to check this observation using real 
experimental data.  

This observation is not only meaningful in the study of jets in e$^+$e$^-$ 
collisions but also enlightening in the jet-physics in relativistic
heavy ion collisions, which will become important~\cite{WXN} after the 
operation of the new generation of heavy ion colliders at BNL 
(RHIC)\cite{Harris} and CERN (LHC-ALICE)\cite{LoI}.

\vskip0.5cm
\n{\bf Acknoledgement} 

\n The authors are grateful to Wu Yuanfang
and Xie Qubin for valuable discussions.

\newpage
\def\Journal#1#2#3#4{{#1} {\bf #2} (#4) #3}
\def\NCA{\em Nuovo Cimento} \def\NIM{\em Nucl. Instrum. Methods}
\def\NIMA{{\em Nucl. Instrum. Methods}| {\bf A}}
\def\NPB{{\em Nucl. Phys.} {\bf B}}
\def\PLB{{\em Phys. Lett.} {\bf B}} \def\PRL{\em Phys. Rev. Lett.}
\def\PRD{{\em Phys. Rev.} {\bf D}} \def\ZPC{{\em Z. Phys.} {\bf C}}

\ed
\begin{thebibliography}{99}

\bibitem{Jade} W. Bartel et al. (JADE COll.), {\em Phys. Lett.}
{\bf B 123} (1983) 460; {\em Z. Phys.} {\bf C 33} (1986) 23.

\bibitem{Durham} Yu. L. Dokshitzer, {\em J. Phys.} {\bf G 17} (1991) 1537.

\bibitem{Dokshitzer} Yu. L. Dokshitzer, G. D. Leder, S. Moretti and 
B. R. Webber, {\em JHEP} {\bf 08} (1997) 001.

\bibitem{Hanson} G. Hanson et al., {\em Phys. Rev. Lett.} {\bf 35}
(1975) 1609.

\bibitem{JELIS} J. Ellis et al., {\em Nucl. Phys.} {\bf B111}
(1976) 253.

\bibitem{Brandelik} R. Brandelik et al. (TASSO Coll.), {\em Phys. Lett.}
{\bf 86 B} (1979) 243. 

\bibitem{Barber} D. P. Barber (Mark J Coll.), {\em Phys. Rev. Lett.}
{\bf 43} (1979) 830.

\bibitem{Berger} Ch. Berger et al. (PLUTO Coll.),  {\em Phys. Lett.}
{\bf 86 B} (1979) 418. 

\bibitem{Bartel} W. Bartel et al. (JADE Coll.),   {\em Phys. Lett.}
{\bf 91 B} (1980) 142.

\bibitem{FFLPRD} Liu Feng, Liu Fuming and Liu Lianshou, 
\Journal{\PRD}{59}{114020}{1999}.

\bibitem{BP}  A. Bia\l as and R. Peschanski,
\Journal{\NPB}{273}{703}{1986}; \Journal{\em ibid}{308}{857}{1988}.

\bibitem{Ochs} W. Ochs, \Journal{\PLB}{347}{101}{1990}.

\bibitem{MMCN} Liu Lianshou, Zhang Yang and Deng Yue,
\Journal{\ZPC}{73}{535}{1997}.

\bibitem{ZGKX} Wu Yuanfang and Liu Lianshou,
\Journal{{\em Science in China} (series A)}{38} {435}{1995}.

\bibitem{NA22}  N. M. Agabayan et al. (NA22),
\Journal{\PLB}{382}{305}{1996}; N. M. Agabayan et al. (NA22),
\Journal{\PLB}{431}{451}{1998}.

\bibitem{WLprl} Wu Yuanfang and Liu Lianshou, \Journal
{\PRL}{21}{3197}{1993}.  

\bibitem{Vineziano}  G. Veneziano, Momentum and colour structure of jet
in QCD, talk given at the {\em 3rd Workshop
on Current Problems in High Energy Particle Theory}, Florence, 1979.

\bibitem{DELPHI}  P. Abreu (DELPHI), {\em Nucl. Phys.} {\bf B386}
(1992) 471.

\bibitem{WXN} X.-N. Wang,  {\em Phys. Reports} {\bf 280} (1997) 287.

\bibitem{Harris} John Harris, {\em Relativistic Heavy Ion Physics and
Relativistic Heavy Ion Collider}, in Proc. of the Lake Louise Winter 
Institute on Quantum Chromodynamics, Feb. 1998, World Scientific, Singapore.

\bibitem{LoI} {\em Letter of Intent for A Large Ion Collider Experiment},
CERN/LHCC/93-16, 1993.

\end{thebibliography}
